\documentclass{aa}  
\usepackage{txfonts,textcomp}
\usepackage{graphicx}
\usepackage[export]{adjustbox}
\usepackage{natbib}
\usepackage{subcaption}
\usepackage{hyperref}
\usepackage{placeins}
\usepackage{gensymb}

\bibpunct{(}{)}{;}{a}{}{,}

\newcommand{\Mjup}{M$_{\rm Jup}$\,}
\newcommand{\Mjupv}{M$_{\rm Jup}$}

\newcommand{\Msun}{M$_{\sun}$}

\newcommand{\Mstar}{M$_{\rm \star}$}

\begin{document} 

   \title{Updated characterization of long-period single companion by combining radial velocity, relative astrometry, and absolute astrometry}

   %\subtitle{}

   \titlerunning{Updated characterisation of long-period single companion}

   \author{F. Philipot\inst{1}\fnmsep\thanks{Please send any requests to florian.philipot@obspm.fr}
    \and
   A.-M. Lagrange\inst{1,} \inst{2}    
        \and
       P. Rubini\inst{3}
        \and
        F. Kiefer \inst{1}
        \and
        A. Chomez \inst{1}
          }

   \institute{
LESIA, Observatoire de Paris, Universit\'{e} PSL, CNRS, 5 Place Jules Janssen, 92190 Meudon, France
\and
Univ. Grenoble Alpes, CNRS, IPAG, 38000 Grenoble, France
\and
Pixyl S.A. La Tronche, France
       }
   \date{Received XXX  / Accepted XXX}

% \abstract{}{}{}{}{} 
% 5 {} token are mandatory
 
  \abstract
   {Thanks to more than 20 years of monitoring, the radial velocity (RV) method has detected long-period companions (P > 10yr) around several dozens of stars. Yet, the true nature of these companions remains unclear because of the uncertainty as to the inclination of the companion orbital plane.}
   {We wish to constrain the orbital inclination and the true mass of long-period single companions.}
   {We used a Markov Chain Monte Carlo (MCMC) fitting algorithm to combine RV measurements with absolute astrometry and, when available, relative astrometry data.}
   {We have lifted the sin(\textit{i}) indetermination for seven long-period companions. We find true masses in the planetary mass range for the candidate planets detected in the following systems: Epsilon Indi A, HD 13931, HD 115954, and HD 222155. The mass of HD 219077 b is close to the deuterium-burning limit and its nature is uncertain because of the imprecise mass of the host star. Using additional RV measurements, we refine the orbital parameters of HIP 70849 b and find a mass in the planetary range. By combining RV data with absolute and relative astrometry, we significantly improve the characterization of HD 211847 B and properly determine its mass, which appears to be in the low-mass star range. This work illustrates how Gaia and Hipparcos allow for the orbital properties and masses of long-period RV companions to be further constrained.} 
   {}


   \maketitle

%
%-------------------------------------------------------------------

\section{Introduction}

In the last decade, several long-period giant planets have been detected using the radial velocity (hereafter RV) method thanks to the increasing temporal baselines of different surveys (\cite{2011arXiv1109.2497M}, \cite{2020MNRAS.492..377W}, \cite{2021ApJS..255....8R}). Yet, a precise determination of the orbital parameters and mass of the planets is very difficult when the orbital period is much larger than the RV time baseline. As a consequence, the radial distribution of planets beyond 8-10 au -- such as those found by \cite{2019ApJ...874...81F} and \cite{2021ApJS..255...14F} based on the results of the two long RV surveys of \cite{2011arXiv1109.2497M} and \cite{2021ApJS..255....8R}, respectively -- are questionable. This unfortunately prevents an accurate comparison with formation model outputs from being made.

Combining RV data with other methods such as relative or absolute astrometry can, in principle, improve the orbital characterization of these companions. Furthermore, it can also remove the uncertainty of the orbital inclination and then allow us to determine the true mass of the planets. 

Coupling RV data with relative astrometry from direct imaging (hereafter DI) or interferometry has been, however, limited to very few cases since high-contrast imaging (hereafter HCI) or interferometry observations favor young systems to minimize the flux contrast between the star and its companion while RV observations favor old and inactive stars which produce low RV jitters. However, when possible, such a coupling is very efficient. An illustration is the HD 7449 system for which the outer companion was first reported as a planet candidate using only RV data (\cite{2011arXiv1109.2497M}, \cite{2019MNRAS.484.5859W}), and it was then identified as a low-mass star by combining RV data with HCI observations \citep{2016ApJ...818..106R}.

In the 2000s, the combination of RV data and absolute astrometry, thanks to the Fine-Guidance-Sensor onboard the Hubble Space Telescope, also allowed for the inclination of a few stellar systems to be constrained and a few candidate planets to be confirmed (\cite{2002AJ....124.1695B}, \cite{2006AJ....132.2206B}), while others were finally identified as brown dwarfs or low-mass stars (\cite{2007AJ....134..749B}; \cite{2010AJ....139.1844B}). Today, the position and proper motion measurements obtained with the telescopes Hipparcos (\cite{1997A&A...323L..49P}, \cite{2007A&A...474..653V}) and Gaia \citep{2020yCat.1350....0G} allow us to combine the RV data and more precise absolute astrometry for a large number of systems. Since the publication of the first Gaia data release (DR1), a few studies have proven the efficiency of combining RV data with absolute and/or relative astrometry to improve the constraints on the orbital parameters and mass of a companion (\cite{2019A&A...629C...1G}, \cite{2019AJ....158..140B}, \cite{2020A&A...642A..31D}, \cite{2020A&A...642A..18L}, \cite{2020AJ....159...71N}, \cite{2021AJ....162...12V}, \cite{2021AJ....161..179B}, \cite{2021AJ....162..301B}, \cite{2021A&A...645A...7K}, \cite{2021AJ....162..266L}, \cite{2022ApJS..262...21F}). \\

In this paper, we focus on seven long-period single companions detected by the RV method, and combine the available RV data with Hipparcos and Gaia early data release 3 (hereafter EDR3) absolute astrometry and, when available, relative astrometry, to improve the orbital parameters and determine the true mass of these companions. In Section 2, we describe our target selection method and present the RV, HCI, and astrometric data used in our study. Section 3 presents the method used to perform the orbital fitting and, there, we provide the new orbital parameters and mass found for each target. Finally, we discuss the results in Section 4.
   
\section{Target selection and data}

\subsection{Target selection}

We first selected the planetary systems in the exoplanet.eu catalog \citep{2011A&A...532A..79S} for which a single companion has been reported with a semi-major axis greater than 5 au using the RV method. Twenty-five companions were found with such criteria. For nine of them (HD 13724 B, HD 25015 b, HD 181234 b, and HD 219828 B \citep{2022ApJS..262...21F}; HD 92987 B  \citep{2021AJ....162...12V}; HIP 36985 B \citep{2022A&A...658A.145B}; and HD 98649 b, HD 196067 b, and HD 221420 B \citep{2021AJ....162..266L}), the orbital parameters and the true mass have already been properly determined in previous studies. We, therefore, do not consider them in the present study.

We first discarded the HD 95872 system because no Hipparcos data were available. We then discarded four systems for which the available RV time series did not cover both extrema of the RV variations and the orbital period could not be properly determined (HD 26161, HD 120066, HD 150706, and HD 213472). In those four cases, the combination of RV and absolute astrometry did not allow us to constrain the orbital parameters, the orbital inclination, or the true mass of the companion. Finally, we discarded four systems for which the orbital period was well covered by the RV data, but the coupling with absolute astrometry did not allow us to constrain the orbital inclination (HD 136925, HD 190984, HD 220773, and HD 238914). Indeed, the variations in position and acceleration of the proper motion of these stars were too small to constrain the orbital inclination of the companion due to the low mass of the companion (< $\sim$2 \Mjupv) and/or the distance of the system.

Thus, we were left with seven systems for which the addition of absolute astrometry and/or new RV measurements and/or relative astrometry measurements allowed us to determine the exact nature of the companion: Epsilon Indi Ab, HD 13931, HD 1159554, HD 211847, HD 219077, HD 222155, and HIP 70849. For three of these companions (Epsilon Indi Ab, HD 211847 B, and HD 219077 b), a first estimation of their orbital inclination and true mass has been obtained by combining RV data and absolute astrometry. Yet, thanks to additional data or more precise astrometric measurements, we obtained more precise and significantly different results from those reported in the previous studies for six of these companions. In the case of HD 219077 b, the differences were mainly found for the mass of the companion. They are mainly due to the uncertainties as to the host star's mass. 

\subsection{RV data}

The RV data used in this study were obtained with different spectrographs between 1997 and 2021. The HARPS \citep{2003Msngr.114...20M} data were taken from the ESO archives; the ELODIE \citep{1996A&AS..119..373B} and SOPHIE \citep{2008SPIE.7014E..0JP} data were retrieved from the OHP archives; and the CORALIE \citep{1999astro.ph.10223Q}, the HIRES \citep{1994SPIE.2198..362V}, the UVES \citep{2000SPIE.4008..534D}, the AAT \citep{1990SPIE.1235..562D}, the CES \citep{1982SPIE..331..232E} long camera (LC), and the CES very long camera (VLC) data were taken from the literature.

As instrument upgrades can lead to new RV offsets, the same instrument before and after a major upgrade is considered as two different instruments. Consequently, HARPS data obtained before and after the optical fiber upgrade in 2015 \citep{2015Msngr.162....9L} are referred to as H03 and H15, respectively. The SOPHIE data obtained before and after the spectrograph upgrade in 2011 \citep{2013A&A...549A..49B} are referred to as SOPHIE and SOPHIE+, respectively. The HIRES data obtained before and after the upgrade of the spectrograph in 2004 \citep{2019MNRAS.484L...8T} are referred to as Hir94 and Hir04, respectively. Finally, the CORALIE spectrograph had two major upgrades in 2007 \citep{2010A&A...511A..45S} and in 2014. The data obtained before 2007 and after 2014 are referred to as C98 and C14, respectively, and the data obtained between 2007 and 2014 are referred to as C07.

\subsection{Direct imaging data}
In three cases, HCI data are available in the SPHERE archive and can provide relative astrometry. 
The three targets were observed in angular (and spectral) differential imaging (A(S)DI, \cite{2006ApJ...641..556M}) using the telescope in pupil tracking mode. The standard observing mode of SPHERE was used, with IRDIS \citep{2008SPIE.7014E..3LD} dual band images at H2 and H3 (K1 and K2, respectively) and IFS \citep{2008SPIE.7014E..3EC} data covering the YJ (YJH, respectively) bands.
The observing log is given in Table.\ref{obs_log}. Whenever possible, the robust PACO A(S)DI algorithm (\cite{2020A&A...637A...9F}, \cite{2020A&A...634A...2F}) was used. The processing step takes advantage of the developments made to the COBREX data center pipeline (the prereduction improvement as well as the improvement of the detection capability of PACO).
If a dataset did not sufficiently cover the field-of-view rotation to apply ADI-based algorithms, the SPECAL 
\citep{2018A&A...615A..92G} No-ADI algorithm was used. 

In those three systems, only one companion was detected (HD 211847 B). The detected companion is characterized in Table.\ref{table_relative_astrometry}. No detection above 5$\sigma$ was found around HD 219077. Six sources were detected around HIP 70849 but, given their position in a color-magnitude diagram and their separations, they are likely background sources.

\begin{table*}[h!]
\centering
\caption{Observing logs.}
\begin{adjustbox}{width=\textwidth}
\begin{tabular}{ccccccccc}
\hline 
STAR & DATE OBS & FILTER & DIT(s)$\times$Nframe$^a$ & $\Delta$ PA ($\degree$)$^a$ & Seeing (")$^b$ & Airmass$^b$ & $\tau_0$ (ms)$^{a,b}$ & Program ID \\ 
\hline 
\hline 
HIP 70849 & 2015-05-05 & DB\_H23 & 64x64 & 40.4 & 1.12 & 1.08 & 0.0012 & 095.C-0298(A) \\ 
\hline 
HD 211847 & 2015-06-10 & DB\_K12 & 64x8 & 5.8 & 1.31 & 1.05 & 0.0025 & 095.C-0476(A) \\ 
\hline 
HD 219077 & 2015-06-09 & DB\_K12 & 8x64 & 3.3 & 1.25 & 1.31 & 0.0026 & 095.C-0476(A) \\ 
\hline 
\vspace{0.1cm}
\end{tabular}
\end{adjustbox}
\textbf{Notes :} $^a$: DIT is the detector integration time per frame. $\Delta$ PA is the amplitude of the parallactic rotation. $\tau_0$ corresponds to the coherence time. $^b$ are values extracted from the updated DIMM information, averaged over the sequence. 
\label{obs_log}
\end{table*}

\begin{table}[h!]
\centering
\caption{Relative astrometry for HD 211847 companion.}
\begin{adjustbox}{width=0.5\textwidth}
\begin{tabular}[h!]{cccccc}
\hline
Sources & JD - 2400000 & IRDIS filter & SEP (mas) & PA ($\deg$) \\ 
\hline
HD 211847 B & 57183.39 & K12 & 220$ \pm $4.73 & 194.5$ \pm $2.23  \\
\hline
\vspace{0.05cm}
\end{tabular}
\end{adjustbox}
\textbf{Notes :} The errors displayed are 1$\sigma$. The relative astrometry combines the astrometry measured in the dual bands.
\label{table_relative_astrometry}
\end{table}

\subsection{Absolute astrometry}
We used measurements from Hipparcos obtained around epoch 1991.25 and from Gaia EDR3 (\cite{2016A&A...595A...1G}, \cite{2021A&A...649A...1G}) obtained around epoch 2016.0.
For each target, the stellar acceleration was determined from the proper motion and the position values were measured by Hipparcos and Gaia with an interval of about 25 years. We considered the proper motion values published by \cite{2021ApJS..254...42B} in the Hipparcos-Gaia Catalog of Accelerations (HGCA). Moreover, a more accurate tangential proper motion ($\mu_{Hip-EDR3}$) was estimated by the difference between the position measurements obtained by Hipparcos and Gaia divided by the time interval between the two measurements ($\sim$25 years). The proper motion values used for each star are given in table \ref{PMvalues}.

\section{Updated orbital parameters and mass}

\subsection{Orbit fitting}

Orbits were fitted using a custom MCMC tool, based on the emcee 3.0 python package \citep{Foreman_Mackey_2013}. It uses a mixture of move functions (such as the differential evolution move function) to alleviate potential multimodality issues. The Hipparcos/Gaia data processing uses the HTOF package \citep{Brandt_2021_htof} and borrows large sections of the orvara code \citep{Brandt_2021_orvara} for the likelihood computation. The HTOF package \citep{2021AJ....162..230B} was used to fit the intermediate astrometric data (IAD) from Hipparcos, based on the 1997 \citep{1997yCat.1239....0E} and 2007 \citep{2007A&A...474..653V} reductions and from Gaia, thanks to the Gaia Observation Forecast Tool (GOST) which allowed us to obtain the estimated Gaia observations and scan angles for each target, in order to reproduce proper motion and position of each observation. Using the Hipparcos and Gaia positions and the temporal baseline, the algorithm derived a tangential proper motion value that allowed us to better constrain the orbital fit when combined with RV data.

We considered ten free parameters for each system: the semi-major axis (\textit{a}), the eccentricity, the orbital inclination (\textit{i}), the host star mass, the companion mass, the longitude of ascending node ($\Omega$), the argument of periastron ($\omega$), the phase, a stellar jitter, and the distance of the system. In addition, to combine data from different instruments, we added an instrumental offset for each instrument as a free parameter of the model (see above). Finally, we considered uniform priors for all fitting parameters, except for the host star mass and the distance of the system for which we considered Gaussian priors.

\subsection{Results}

\subsubsection{Epsilon Indi A}

Epsilon Indi is a triple system with a $0.76 \pm 0.04$ \Msun, K2V star (Epsilon Indi A) and a binary composed of a $75.0 \pm 0.8$ \Mjupv, T1.5 brown dwarf (Epsilon Indi B) and a $70.1 \pm 0.7$ \Mjupv, T6 brown dwarf (Epsilon Indi C) separated by about 2.6 au \citep{2018ApJ...865...28D}. The projected separation between the binary brown dwarfs and the star is about 1460 au. Combining RV data and absolute astrometry based on Hipparcos and the Gaia data release 2 (DR2) measurements, \cite{2019MNRAS.490.5002F} reported a giant planet with a semi-major axis of $11.55^{+0.98}_{-0.86}$ au, a mass of $3.25^{+0.39}_{-0.65}$ \Mjupv, an inclination of $64.25^{+13.80}_{-6.09}$°, and an eccentricity of $0.26^{+0.07}_{-0.03}$. Yet, the Gaia EDR3 proper motion and position measurements are significantly more precise compared to the Gaia DR2 measurement and they significantly improve the characterization of Epsilon Indi Ab. 

\begin{figure}[t!]
  \centering
\includegraphics[width=0.45\textwidth]{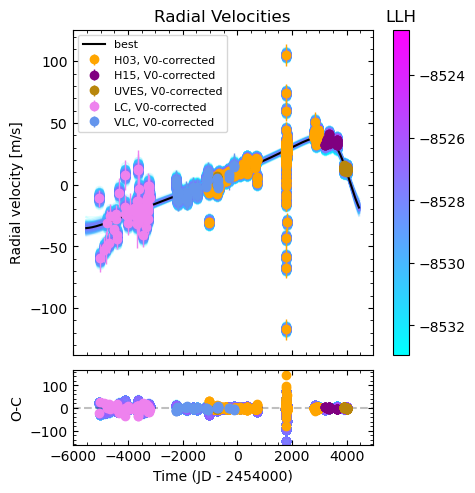}
\includegraphics[width=0.5\textwidth]{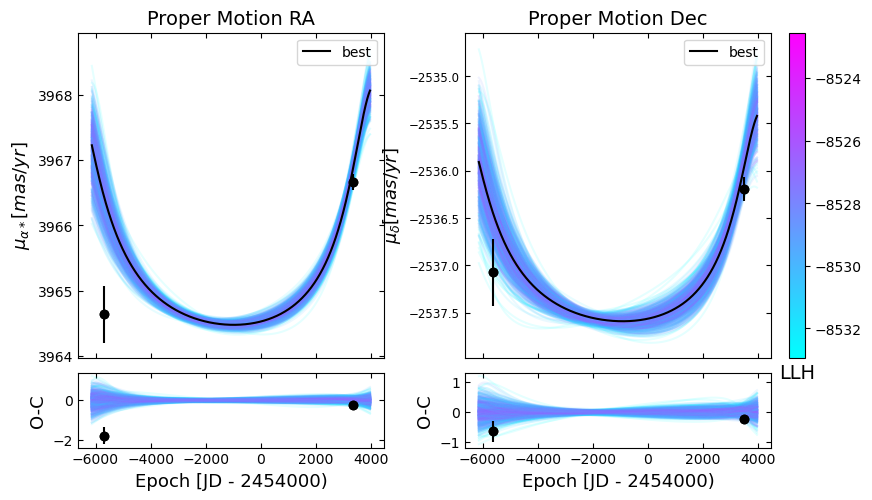}
\caption{Orbital fits for Epsilon Indi Ab. \textit{Top}: Fit of the Epsilon Indi A RV data corrected from the instrumental offset (V0). \textit{Bottom}: Fit of the Epsilon Indi A astrometric acceleration in right ascension (left) and declination (right). The black points correspond to the measurements obtained with Hipparcos (1991.25) and Gaia EDR3 (2016.0). In each plot, the black curve shows the best fit. The color bar indicates the log likelihood of the different fits plotted.
\label{RV_AA_EpsInd}} 
\end{figure}

We used 4278 RV measurements obtained with the HARPS spectrograph between 2003 and 2016\footnote{The 3636 RV data obtained between Julian days 2455790 and 2455805 were obtained to study high-frequency oscillations of the star. These data were measured with high cadence, which led to a significantly lower signal-to-noise ratio compared to the other data.}, 163 RV measurements obtained with the UVES spectrograph between 1996 and 2017, 72 RV measurements obtained with the LC spectrograph between 1992 and 1997, and 53 RV measurements obtained with the VLC spectrograph between 2000 and 2006. We also combined these RV data with absolute astrometry based on Hipparcos and the Gaia EDR3 measurements (Fig.\ref{RV_AA_EpsInd}). We found significantly different orbital parameters with a semi-major axis of $8.8^{+0.2}_{-0.1}$ au, a mass of $3.0 \pm 0.1$ \Mjupv, an inclination of $91^{+4}_{-5}$°, and an eccentricity of $0.48 \pm 0.01$. It is important to note that if we consider only the 539 HARPS RV data with a signal-to-noise ratio greater than 110 and thus remove the high cadence observations made in August 2011, we find similar solutions.

\subsubsection{HD 13931}

HD 13931 is a $1.02 \pm 0.05$ \Msun \citep{2021ApJS..255....8R}, G0V star. Based on 66 RV measurements obtained with the HIRES spectrograph between 1998 and 2019, \cite{2021ApJS..255....8R} reported a giant planet with a semi-major axis of $5.323 \pm 0.091$ au, a minimum mass of $1.911^{+0.077}_{-0.076}$ \Mjup, and an eccentricity of $0.02^{+0.021}_{-0.014}$.

\begin{figure}[t!]
  \centering
\includegraphics[width=0.45\textwidth]{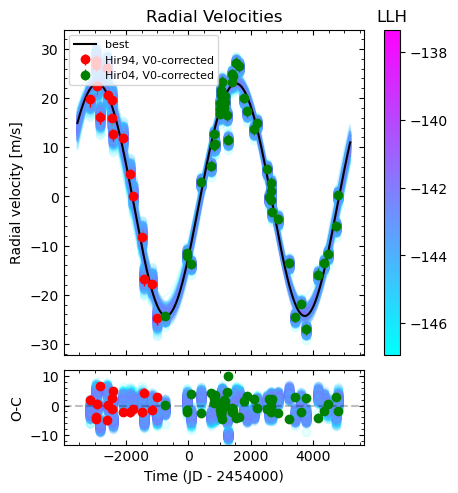}
\includegraphics[width=0.5\textwidth]{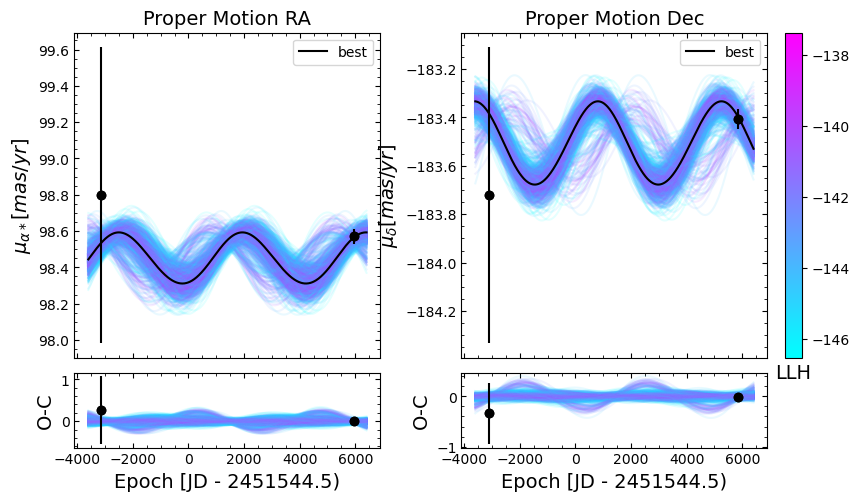}
\caption{Orbital fits for HD 13931 b. \textit{Top}: Fit of the HD 13931 RV data corrected from the instrumental offset (V0). \textit{Bottom}: Fit of the HD 13931 astrometric acceleration in right ascension (left) and declination (right). The black points correspond to the measurements obtained with Hipparcos (1991.25) and Gaia (2016.0). In each plot, the black curve shows the best fit. The color bar indicates the log likelihood of the different fits plotted.
\label{RV_AA_HD13931}} 
\end{figure}

We combined these RV data with absolute astrometry (Fig.\ref{RV_AA_HD13931}). As the RV baseline is much larger than the orbital period, the orbital parameters are well-constrained. As expected, we found a semi-major axis and an eccentricity very close to those reported by \cite{2021ApJS..255....8R} with \textit{a} = $5.33 \pm 0.09$ au and e < 0.04. 
Using, in addition, the absolute astrometry, we found an orbital inclination of either $39^{+13}_{-8}$° or $141^{+9}_{-18}$° and a true mass of $3.1^{+0.8}_{-0.7}$ \Mjupv.

\subsubsection{HD 115954}

HD 115954 is a $1.18 \pm 0.06$ \Msun, G0V star \citep{2021A&A...653A..78D}. Based on four RV measurements obtained with the ELODIE spectrograph between 2004 and 2005 and 45 RV measurements obtained with the SOPHIE spectrograph between 2009 and 2018, \cite{2021A&A...653A..78D} reported a giant planet with a semi-major axis of $5.00^{+1.3}_{-0.36}$ au, a minimum mass of $8.29^{+0.75}_{-0.58}$ \Mjup, and an eccentricity of $0.487^{+0.095}_{-0.041}$. 

\begin{figure}[t!]
  \centering
\includegraphics[width=0.45\textwidth]{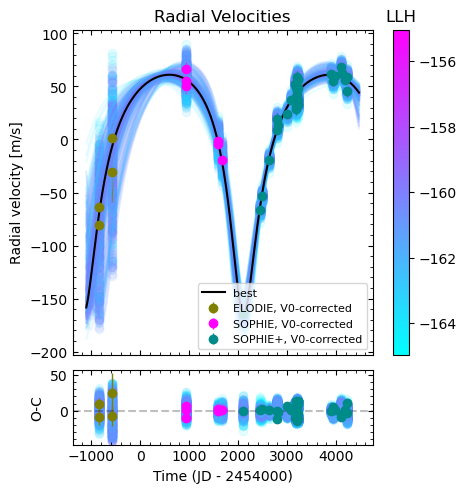}
\includegraphics[width=0.5\textwidth]{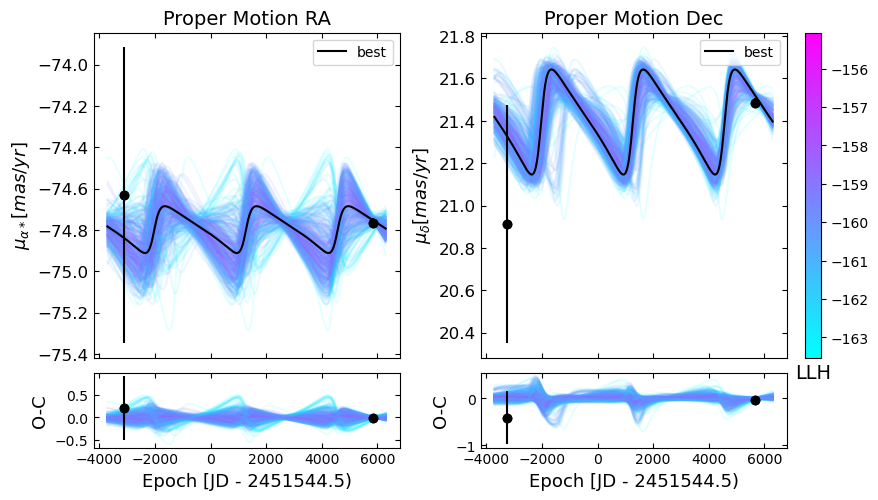}
\caption{Orbital fits for HD 115954 b. \textit{Top}: Fit of the HD 115954 RV data corrected from the instrumental offset (V0). \textit{Bottom}: Fit of the HD 115954 astrometric acceleration in right ascension (left) and declination (right). The black points correspond to the measurements obtained with Hipparcos (1991.25) and Gaia (2016.0). In each plot, the black curve shows the best fit. The color bar indicates the log likelihood of the different fits plotted.
\label{RV_AA_HD115954}} 
\end{figure}

We combined these RV data with the absolute astrometry data (Fig.\ref{RV_AA_HD115954}). We found a semi-major axis compatible with \cite{2021A&A...653A..78D} with \textit{a} = $4.5^{+0.2}_{-0.1}$ au and an eccentricity of $0.46 \pm 0.03$. Using, in addition, the absolute astrometry, we found an orbital inclination of $92^{+17}_{-16}$° and a true mass of $8.5^{+0.6}_{-0.4}$ \Mjupv.

\subsubsection{HD 211847}

HD 211847 is a $0.94 \pm 0.04$ \Msun, G5V star \citep{2011A&A...525A..95S}. \cite{2011A&A...525A..95S} reported a brown dwarf candidate orbiting around HD 211847 based on 31 RV measurements obtained with the CORALIE spectrograph between 2002 and 2009. However, only one minimum of the HD 211847 B RV curve was covered by the dataset. Thus, the orbital parameters and minimum mass reported in this study are poorly constrained. Using the Levenberg-Marquardt method, they found ranges corresponding to a $3\sigma$ confidence interval for the semi-major axis, the eccentricity, and the minimum mass of 4.6--42 au, 0.48--0.95, and 16.3--24.3 \Mjup, respectively. \cite{2017A&A...602A..87M} obtained one HCI detection with SPHERE of HD 211847 B for a projected separation of 11.3 au. Using the BT-Settl models \citep{2014IAUS..299..271A}, they fit the HD 211847 B spectrum and found a low stellar mass of $155 \pm 9$ \Mjup assuming an age of 3 Gyr for the host star. Based on the result of \cite{2011A&A...525A..95S}, \cite{2017A&A...602A..87M} estimated the inclination of the companion orbit to be around seven°. Recently, combining the CORALIE RV measurement and the absolute astrometry, \cite{2022ApJS..262...21F} reported HD 211847 B as a brown dwarf with a semi-major axis of $4.514^{+0.458}_{-0.287}$ au, a mass of $55.32^{+1.335}_{-18.48}$ \Mjupv, an inclination of $163.649^{+36.239}_{-5.017}$°, and an eccentricity of $0.419^{+0.035}_{-0.064}$.

\begin{figure}[t!]
  \centering
\includegraphics[width=0.4\textwidth]{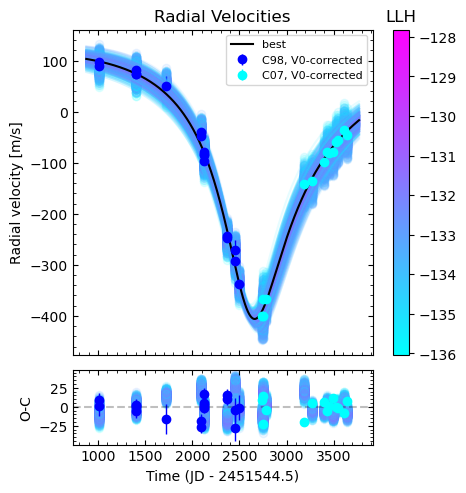}
\includegraphics[width=0.4\textwidth]{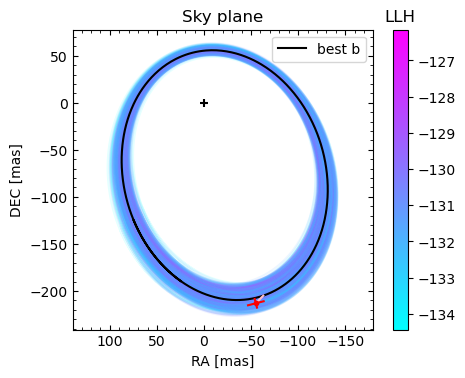}
\includegraphics[width=0.5\textwidth]{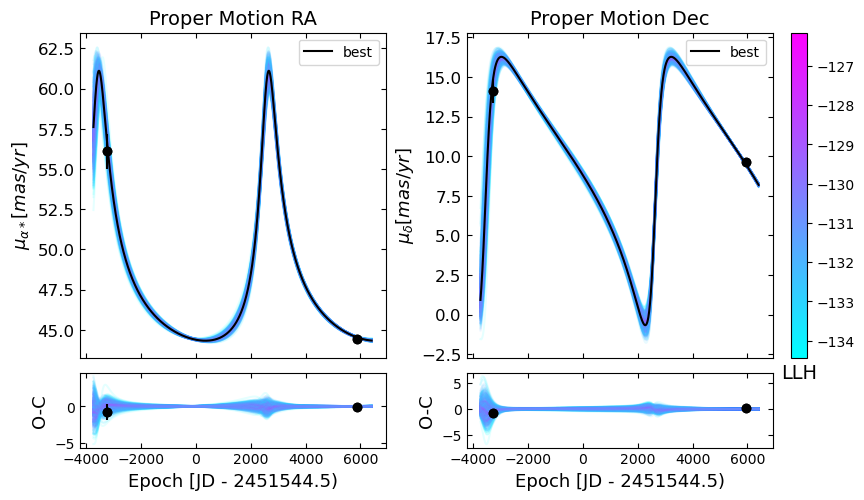}
\caption{Orbital fits for HD 211847 B. \textit{Top left}: Fit of the HD 211847 RV data corrected from the instrumental offset (V0). \textit{Top right}: Fit of HD 211847 relative astrometry data. The red cross corresponds to the measurement obtained with SPHERE. \textit{Bottom}: Fit of the HD 211847 astrometric acceleration in right ascension (left) and declination (right). The black points correspond to the measurements obtained with Hipparcos (1991.25) and Gaia (2016.0). In each plot, the black curve shows the best fit. The color bar indicates the log likelihood of the different fits plotted.
\label{RV_AA_HD211847}} 
\end{figure}

We combined the RV dataset used by \cite{2011A&A...525A..95S}, the relative astrometry observation obtained with SPHERE in June 2015, and the absolute astrometry (Fig.\ref{RV_AA_HD211847}). Adding one relative astrometry observation allowed us to properly constrain the orbital parameters and the mass of HD211847 B with results significantly different from those reported by \cite{2022ApJS..262...21F}. We found a semi-major axis of $6.78 \pm 0.08$ au and an eccentricity of $0.59^{+0.01}_{-0.02}$. Using, in addition, the absolute astrometry, we found an orbital inclination of $172.3^{+0.05}_{-0.04}$° and a true mass of $148 \pm 5$ \Mjupv. We note that by taking only RV data and absolute astrometry into account, we found very poorly constrained solutions with large uncertainties as to the semi-major axis (16 - 30 au) and mass (80 - 140 \Mjupv). Moreover, the solutions found are not in agreement with those reported by \cite{2022ApJS..262...21F} or with the solutions found when adding the HCI data.

\subsubsection{HD 219077}

Based on 63 CORALIE RV measurements obtained between 1999 and 2012 and 30 HARPS RV measurements obtained between 2003 and 2012, \cite{2013A&A...551A..90M} reported a very eccentric giant planet with a semi-major axis of $6.22 \pm 0.09$ au and a minimum mass of $10.39 \pm 0.09$ \Mjupv. It is important to note that the RV data used by \cite{2013A&A...551A..90M} are not publicly available. Based on 72 pieces of RV data obtained with the AAT spectrograph between 1998 and 2015, \cite{2019AJ....157..252K} reported slightly different properties for HD 219077 with a semi-major axis of $7.03^{+0.20}_{-0.21}$ au and a minimum mass of $13.40^{+0.76}_{-0.78}$ \Mjupv. These differences are probably due to the different assumptions on the mass of the star. Indeed, \cite{2013A&A...551A..90M} used the values reported by Hipparcos (\Mstar = $1.05 \pm 0.02$ \Msun ) while \cite{2019AJ....157..252K} used the values reported in \cite{2005yCat..21590141V} (\Mstar = $1.51 \pm 0.13$ \Msun ). Recently, \cite{2022ApJS..262...21F} combined the AAT RV measurements used by \cite{2019AJ....157..252K} and 33 HARPS RV measurements obtained between 2003 and 2012 with the absolute astrometry based on Hipparcos and the Gaia EDR3 measurements and found a semi-major axis close to \cite{2013A&A...551A..90M}, an orbital inclination of $90.178^{+9.527}_{-9.462}$°, and a true mass of $9.620^{+1.001}_{-0.733}$ \Mjupv. The prior on the mass of the star is not given.

\begin{figure}[t!]
  \centering
\includegraphics[width=0.45\textwidth]{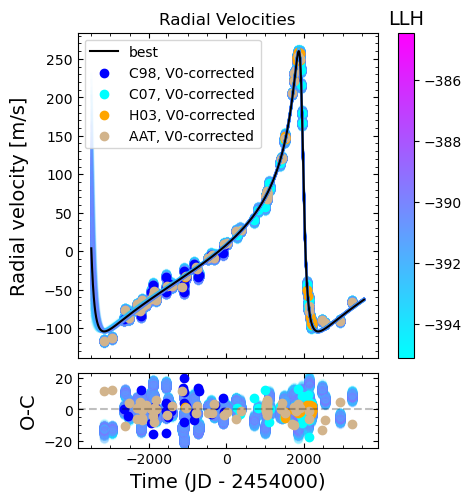}
\includegraphics[width=0.5\textwidth]{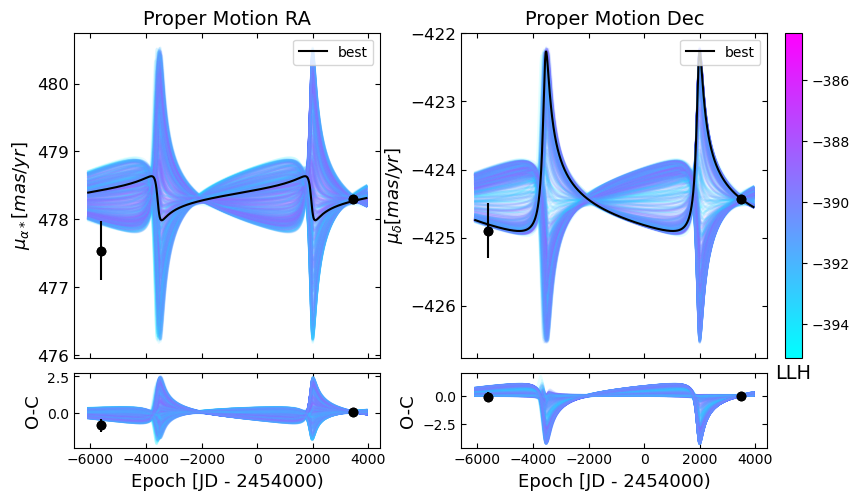}
\caption{Orbital fits for HD 219077 b. \textit{Top}: Fit of the HD 219077 RV data corrected from the instrumental offset (V0). \textit{Bottom}: Fit of the HD 219077 astrometric acceleration in right ascension (left) and declination (right). The black points correspond to the measurements obtained with Hipparcos (1991.25) and Gaia (2016.0). In each plot, the black curve shows the best fit. The color bar indicates the log likelihood of the different fits plotted.
\label{RV_AA_HD219077}} 
\end{figure}

For this study, as the data used by \cite{2013A&A...551A..90M} are not available, we considered the HARPS and AAT RV measurements used on \cite{2022ApJS..262...21F} and the 65 CORALIE RV measurements available on the DACE archive\footnote{https://dace.unige.ch} obtained between 1999 and 2012\footnote{The CORALIE RV data available on DACE and the HARPS RV data available on the ESO archive cover the same time base as those used by \cite{2013A&A...551A..90M}.}. For the mass of the star, we considered the value given by \cite{2022A&A...657A...7K} based on the Gaia DR3 results (\Mstar = $1.15 \pm 0.06$ \Msun). We combined the RV data with the absolute astrometry (Fig.\ref{RV_AA_HD219077}). We found a semi-major axis and an eccentricity close to those reported in the previous studies with \textit{a} = $6.4 \pm 0.1$ au and e = $0.769 \pm 0.002$ and an orbital inclination close to that of \cite{2022ApJS..262...21F} with either \textit{i} = $83 \pm 3$° or \textit{i} = $97 \pm 3$°. Considering the star mass found by \cite{2022A&A...657A...7K}, we found a planetary mass at $11.3 \pm 0.4$ \Mjupv. However, considering that the mass used in \cite{2005yCat..21590141V} would lead to a mass close to the deuterium-burning limit, $M_{b}$ = $13.6 \pm 0.5$ \Mjupv. Due to the uncertainties on the mass of the host star, it is not possible to determine the exact nature of HD 219077 b.

\subsubsection{HD 222155}

HD 222155 is a $1.13 \pm 0.11$ \Msun, G2V star \citep{2012A&A...545A..55B}. Based on 44 RV measurements obtained with the ELODIE spectrograph between 1997 and 2005 and 67 RV measurements obtained with the SOPHIE spectrograph between 2007 and 2011, \cite{2012A&A...545A..55B} reported a giant planet with a semi-major axis of $5.1^{+0.6}_{-0.7}$ au, a minimum mass of $1.90^{+0.67}_{-0.53}$ \Mjup, and an eccentricity of $0.38^{+0.28}_{-0.32}$.

\begin{figure}[t!]
  \centering
\includegraphics[width=0.45\textwidth]{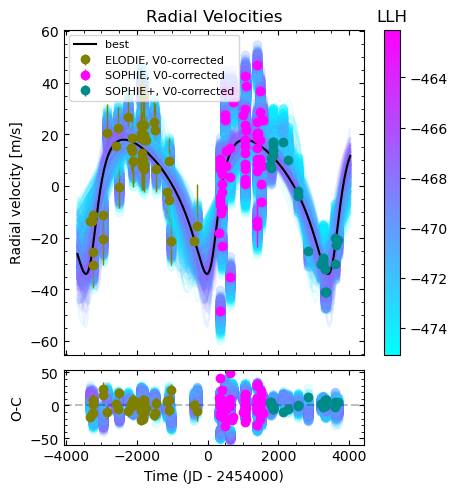}
\includegraphics[width=0.5\textwidth]{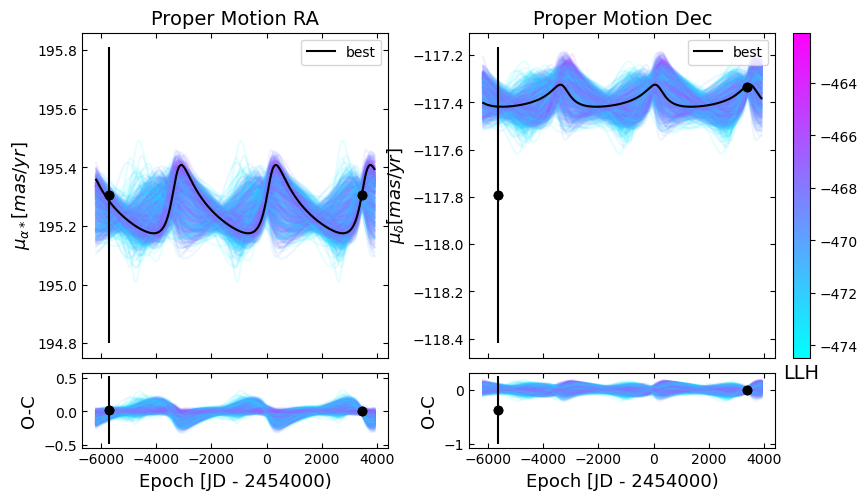}
\caption{Orbital fits for HD 222155 b. \textit{Top}: Fit of the HD 222155 RV data corrected from the instrumental offset (V0). \textit{Bottom}: Fit of the HD 222155 astrometric acceleration in right ascension (left) and declination (right). The black points correspond to the measurements obtained with Hipparcos (1991.25) and Gaia (2016.0). In each plot, the black curve shows the best fit. The color bar indicates the log likelihood of the different fits plotted.
\label{RV_AA_HD222155}} 
\end{figure}

We considered 31 additional pieces of SOPHIE RV data obtained between 2011 and 2016. We combined the RV data with the absolute astrometry (Fig.\ref{RV_AA_HD222155}). We found orbital parameters within the error bars associated with the values found by \cite{2012A&A...545A..55B} with \textit{a} = $4.7 \pm 0.1$ au and e = $0.34 \pm 0.09$. As the RV baseline is now much larger than the orbital period, the orbital parameters are better constrained. Using, in addition, the absolute astrometry, we found an orbital inclination of either $66^{+14}_{-11}$° or $115^{+13}_{-16}$° and a true mass of $2.1^{+0.3}_{-0.2}$ \Mjupv.

\subsubsection{HIP 70849}

HIP 70848 is a $0.63 \pm 0.03$ \Msun, K7V star \citep{2011A&A...535A..54S}. \cite{2011A&A...535A..54S} reported the first detection of HIP 70849 b based on 18 RV measurements obtained with the HARPS spectrograph between 2006 and 2010. However, only one minimum of the HIP 70849 b RV curve was covered by the dataset. The observations carried out by \cite{2011A&A...535A..54S} led to poorly constrained orbital parameters and minimum mass. Using a genetic algorithm followed by MCMC simulations, they reported a semi-major axis between 4.5 and 36 au, a minimum mass between 3 and 15 \Mjup, and an eccentricity between 0.47 and 0.96 with ranges corresponding to a $3\sigma$ confidence interval.

\begin{figure}[t!]
  \centering
\includegraphics[width=0.45\textwidth]{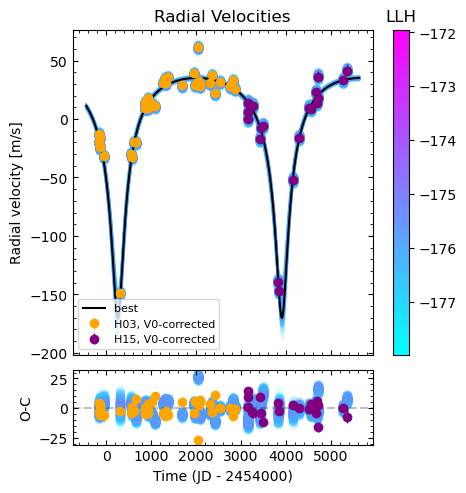}
\includegraphics[width=0.5\textwidth]{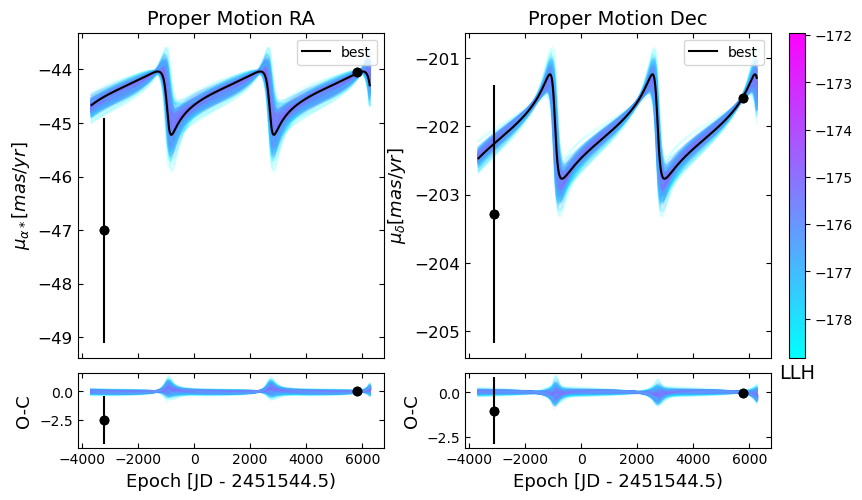}
\caption{Orbital fits for HIP 70849 b. \textit{Top}: Fit of the HIP 70849 RV data corrected from the instrumental offset (V0). \textit{Bottom}: Fit of the HIP 70849 astrometric acceleration in right ascension (left) and declination (right). The black points correspond to the measurements obtained with Hipparcos (1991.25) and Gaia (2016.0). In each plot, the black curve shows the best fit. The color bar indicates the log likelihood of the different fits plotted.
\label{RV_AA_HIP70849}} 
\end{figure}

We considered 39 additional pieces of HARPS RV data obtained between 2011 and 2021. We combined the RV data with the absolute astrometry (Fig.\ref{RV_AA_HIP70849}). With these additional observations, the dataset then covered two minimum and one maximum of the RV curve of HIP 70849 b and this allowed us to properly constrain the properties of the companion. We found a semi-major axis of $3.99^{+0.06}_{-0.07}$ au and an eccentricity of $0.65^{+0.02}_{-0.01}$. Using, in addition, the absolute astrometry, we found an orbital inclination of $96 \pm 16$° and a true mass of $4.5^{+0.4}_{-0.3}$ \Mjupv.

\begin{table*}[h!]
\centering
\caption{Summary of posteriors obtained with our MCMC algorithm.}
\begin{adjustbox}{width=\textwidth}
\begin{tabular}[h!]{ccccccccccc}
\hline
 Parameter & Eps Ind A & HD 13931 & HD 115954 & HD 211847 & HD 219077 & HD 222155 & HIP 70849 \\ 
\hline
 \textit{a} (au) & ${8.8}_{-0.1}^{+0.2}$ & $5.33 \pm 0.09$ & ${4.5}_{-0.1}^{+0.2}$ & $6.78 \pm 0.08$ & $6.4 \pm 0.1$ & $4.7 \pm 0.1$ & ${3.99}_{-0.07}^{+0.06}$ \\ 
 Period (days) & ${10932}_{-228}^{+266}$ & ${4442}_{-46}^{+49}$ & ${3258}_{-190}^{+179}$ & ${6199}_{-46}^{+52}$ & ${5514}_{-39}^{+44} $ & ${3470}_{-106}^{+102}$ & $3649 \pm 18$ \\
 Eccentricity & $0.48 \pm 0.01$ & < 0.04 & $0.46 \pm 0.03$ & $0.769 \pm 0.002$ & ${0.59}_{-0.02}^{+0.01}$ & $0.34 \pm 0.09$ & ${0.65}_{-0.01}^{+0.02}$ \\ 
 Inclination (°) & $91^{+4}_{-5}$ & $39^{+13}_{-8}$ or $141^{+9}_{-18}$ & ${59}_{-4}^{+5}$ or $127 \pm 4$ & ${172.3}_{-0.4}^{+0.5}$ & $83 \pm 3$ or $97 \pm 3$ & ${66}_{-11}^{+14}$ or ${115}_{-16}^{+13}$ & $96 \pm 16$ \\
 Mass (\Mjup) & $3.0 \pm 0.1$ & $3.1^{+0.8}_{-0.7}$ & ${8.5}_{-0.4}^{+0.6}$ & $148 \pm 5$ & $11.3 \pm 0.4$ & ${2.1}^{+0.3}_{-0.2}$ & ${4.5}_{-0.3}^{+0.4}$ \\  
 $\Omega$ (°) & $58 \pm 5$ & ${343}_{-19}^{+17}$ or ${110}_{-24}^{+19}$ & ${211}_{-28}^{+25}$ & $184 \pm 5$ & ${135}_{-21}^{+38}$ or ${347}_{-30}^{+28}$ & ${264}_{-33}^{+34}$ or ${180}_{-35}^{+34}$ & $35 \pm 6$ \\ 
 $\omega$ (°) & $85 \pm 3$ & 74 - 227 & ${173}_{-8}^{+7}$ & ${168}_{-4}^{+5}$ & $56.2 \pm 0.4$ & 153 - 217 & $182 \pm 1$ \\  
 Phase & $0.37 \pm 0.01$ & 0.38 - 0.86 & $0.40_{-0.06}^{+0.09}$ & $0.420 \pm 0.004$ & $0.354_{-0.003}^{+0.002}$ & $0.98_{-0.13}^{+0.04}$ & $0.745_{-0.006}^{+0.007}$ \\ 
 Jitter (m/s) & $3.37 \pm 0.04$ & ${2.9} \pm 0.3$ & $6.6_{-1.1}^{+1.3}$ & ${11.0}_{-2.0}^{+2.8}$ & ${4.3}_{-0.2}^{+0.3}$ & ${12.7}_{-0.8}^{+0.9}$ & ${6.7}_{-0.7}^{+0.8}$ \\
\hline
 & H03 = $-39972 \pm 1$ & Hir94 = $-13 \pm 1$ & ELODIE = ${-14757}_{-41}^{+52}$ & C98 = $6907 \pm 17$ & C98 = ${-30867}_{-1}^{+2}$ & ELODIE = $-3999 \pm 3$ & H03 = $53 \pm 1$ \\
 Instrumental & H15 = ${-39952}_{-1}^{+2}$ & Hir04 = $-8.0 \pm 0.5$ & SOPHIE = ${-14768}_{-8}^{+9}$ & C07 = $6849 \pm 22$ & C07 = ${-30864}_{-9}^{+8}$ & SOPHIE = ${-3950}_{-2}^{+3}$ & H15 = ${65}_{-1}^{+2}$ \\
 offset (m/s) & UVES = $-5 \pm 1$ & & SOPHIE+ = ${-14743}_{-3}^{+4}$ &  & H03 = ${-30830}_{-2}^{+1}$ & SOPHIE+ = ${-3923}_{-3}^{+4}$ & \\
 & LC = ${-39978}_{-1}^{+2}$ & & & & AAT = $-68 \pm 1$ & & \\
 & VLC = ${-39976}_{-1}^{+2}$ & & & & & & \\
\hline
\vspace{0.05cm}
\end{tabular}
\end{adjustbox}
\textbf{Notes:} The results were obtained by combining RV, absolute astrometry, and, when available, relative astrometry. We provide $68\%$ confidence intervals for each parameter and the median is only given when the probability distribution has a profile close to a Gaussian distribution.
\label{table_summary}
\end{table*}

\section{Summary and concluding remarks}

\begin{figure}[t!]
  \centering
\includegraphics[width=0.5\textwidth]{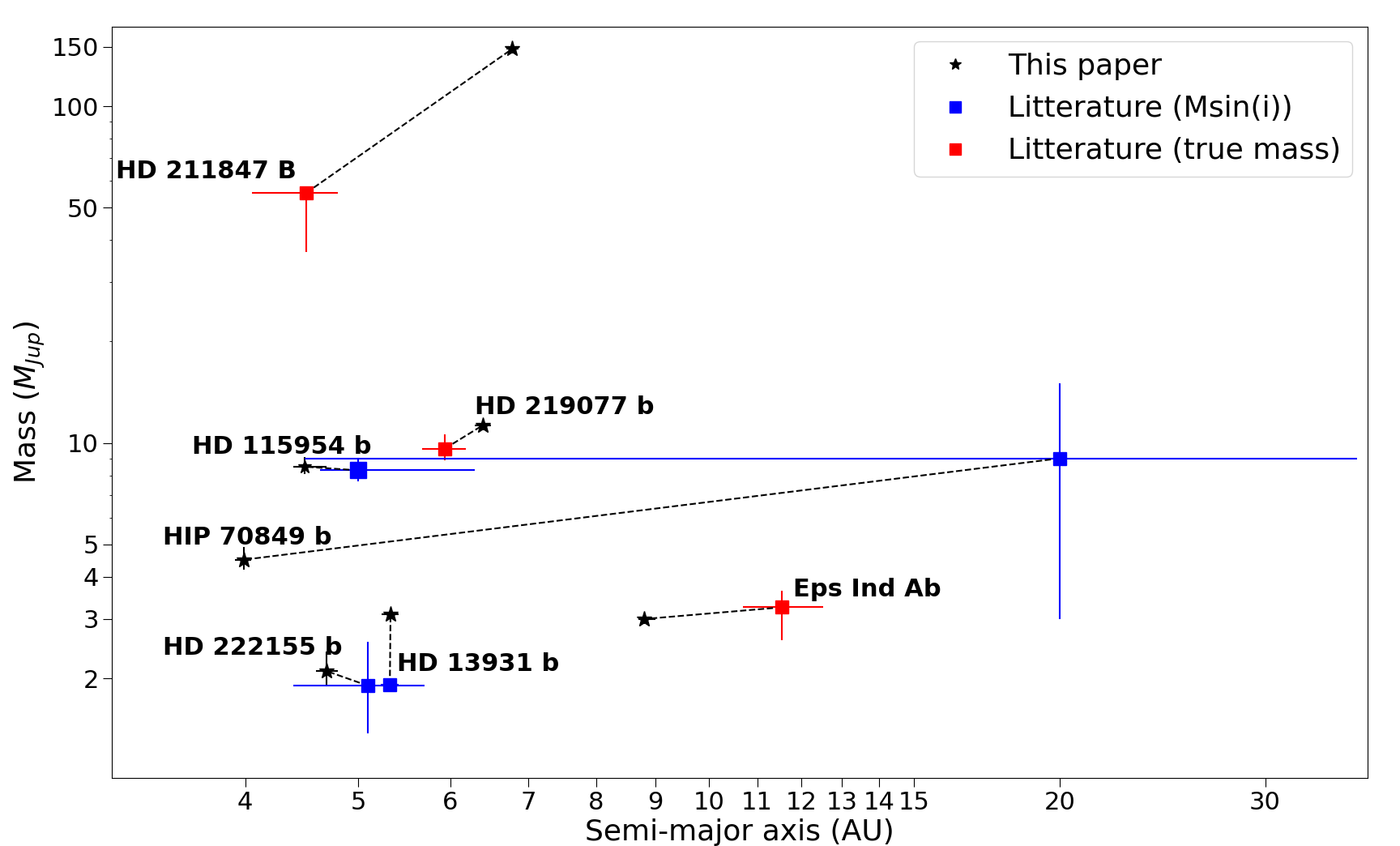}
\caption{Update of the orbital parameters and masses of the seven analyzed systems thanks to the combination of absolute astrometric and RV data and, when available, absolute astrometry data.
For each system, a dotted line between two solutions was drawn to allow for the different solutions obtained to be compared.
\label{sma_mass}} 
\end{figure}

Combining RV measurements from various spectrographs with absolute astrometry based on Hipparcos and Gaia EDR3 data and, when available, relative astrometry, we determined the orbital parameters and, in particular, the orbital inclination and the true mass of seven long-period single companions detected by the RV method. Figure \ref{sma_mass} summarizes the true mass and the semi-major axis of the companions and compares them with the previous estimations. Clearly, Gaia EDR3 data allow for a better determination of these companions' orbital parameters and mass. All of these companions have true masses of 2 \Mjup or more and orbit between 3.9 - 9 au from their stars. Absolute astrometry would probably help to determine the true mass of planets with a period larger than the duration of Gaia DR3 observations (P > $\sim$1000 d) down to 1 \Mjup, provided the RV variations are well covered and the variations in position and acceleration of the proper motion of the star are large enough. In practice, in most cases, when the period is not well constrained by the RV data, the impact of the coupling of RV data with absolute astrometry is more limited. An illustration of this is the case of HD 211847 B for which, by combining RV data that cover only a minimum of the RV variations with absolute astrometry, \cite{2022ApJS..262...21F} reported a mass of about 55 \Mjupv, corresponding to a brown dwarf. Yet, HCI revealed a companion and the fit of the RV and the relative and absolute astrometry leads to a mass of about 150 \Mjup instead.

We conclude that Gaia/Hipparcos can help to further constrain the orbital parameters of long-period RV planets, providing good coverage of the RV variations is available. Otherwise, additional information is needed, such as relative astrometry, provided by DI or interferometry.

\begin{acknowledgements}
      This study was funded by a grant from PSL/OCAV. 
      This project has also received funding from the European Research Council (ERC) under the European Union's Horizon 2020 research and innovation programme (COBREX; grant agreement n° 885593).
      This work presents results from the European Space Agency (ESA) space mission Gaia. Gaia data are being processed by the Gaia Data Processing and Analysis Consortium (DPAC). Funding for the DPAC is provided by national institutions, in particular the institutions participating in the Gaia MultiLateral Agreement (MLA). The Gaia mission website is https://www.cosmos.esa.int/gaia. The Gaia archive website is https://archives.esac.esa.int/gaia.
      This publications makes use of the The Data \& Analysis Center for Exoplanets (DACE), which is a facility based at the University of Geneva (CH) dedicated to extrasolar planets data visualisation, exchange and analysis. DACE is a platform of the Swiss National Centre of Competence in Research (NCCR) PlanetS, federating the Swiss expertise in Exoplanet research. The DACE platform is available at https://dace.unige.ch.
      This research has made use of the SIMBAD database and VizieR catalogue access tool, operated at CDS, Strasbourg, France.
      Based on data retrieved from the SOPHIE archive at Observatoire de Haute-Provence (OHP), available at atlas.obs-hp.fr/sophie.
      Based on spectral data retrieved from the ELODIE archive at Observatoire de Haute-Provence (OHP).
      Based on observations collected at the European Southern Observatory under ESO programme(s) 183.C-0972(A), 108.22KV.001, 108.22KV.002, 072.C-0488(E), 108.222V.001, 090.C-0421(A), 085.C-0019(A), 087.C-0831(A), 095.C-0551(A), 0100.C-0097(A), 0103.C-0432(A), 0101.C-0379(A), 096.C-0460(A), 0102.C-0558(A), 093.C-0409(A), 098.C-0366(A), 099.C-0458(A), 089.C-0732(A), 091.C-0034(A), 196.C-1006(A), 0102.C-0558(A), 106.21R4.001, 183.C-0972(A), 075.C-0332(A), 192.C-0852(A), 190.C-0027(A), 091.C-0936(A), 085.C-0063(A), 086.C-0284(A), 076.C-0155(A), 077.C-0101(A), 082.C-0212(A).
      
\end{acknowledgements}

\bibliographystyle{aa}
\bibliography{CB}

\onecolumn

\begin{appendix}

\section{Proper motion values}

\begin{table*}[h!]
\centering
\caption{Proper motion values from HGCA.}
\begin{adjustbox}{width=\textwidth}
\begin{tabular}[h!]{ccccccccccc}
\hline
 Star & Eps ind A & HD 13931 & HD 115954 & HD 211847 & HD 219077 & HD 222155 & HIP 70849 \\ 
\hline
 $\mu_{Hip}^{\alpha}$ (mas/yr) & $3964.6 \pm 0.4$ & $98.8 \pm 0.8$ & $-74.6 \pm 0.7$ & $56.1 \pm 1.1$ & $477.5 \pm 0.4$ & $195.3 \pm 0.5$ & $-47.0 \pm 2.1$ \\
 $\mu_{Hip}^{\delta}$ (mas/yr) & $-2537.1 \pm 0.4$ & $-183.7 \pm 0.6$ & $20.9 \pm 0.6$ & $14.1 \pm 0.8$ & $-424.9 \pm 0.4$ & $-117.8 \pm 0.6$ & $-203.3 \pm 1.9$ \\ 
 $\mu_{EDR3}^{\alpha}$ (mas/yr) & $3966.7 \pm 0.1$ & $98.57 \pm 0.04$  & $-74.77 \pm 0.02$ & $44.43 \pm 0.03$ & $478.30 \pm 0.03$ & $195.31 \pm 0.02$ & $-44.05 \pm 0.02$ \\
 $\mu_{EDR3}^{\delta}$ (mas/yr) & $-2536.2 \pm 0.1$ & $-183.41 \pm 0.04$ & $21.49 \pm 0.02$ & $9.66 \pm 0.04$ & $-424.43 \pm 0.04$ & $-117.34 \pm 0.02$ & $-201.58 \pm 0.3$ \\  
 $\mu_{Hip-EDR3}^{\alpha}$ (mas/yr) & $3965.02 \pm 0.01$ & $98.45 \pm 0.03$ & $-74.79 \pm 0.02$ & $47.90 \pm 0.04$ & $478.36 \pm 0.01$ & $195.25 \pm 0.02$ & $-44.43 \pm 0.06$ \\ 
 $\mu_{Hip-EDR3}^{\delta}$ (mas/yr) & $-2537.25 \pm 0.01$ & $-183.51 \pm 0.02$ & $-21.41 \pm 0.02$ & $-10.44 \pm 0.03$ & $-424.40 \pm 0.01$ & $-117.39 \pm 0.02$ & $-202.05 \pm 0.04$ \\
\hline
\vspace{0.05cm}
\end{tabular}
\end{adjustbox}
\textbf{Notes:} $\mu_{Hip}$ corresponds to the proper motion obtained by Hipparcos. $\mu_{EDR3}$ corresponds to the proper motion obtained by Gaia EDR3. $\mu_{Hip-EDR3}$ corresponds to the proper motion obtained by the Hipparcos-Gaia EDR3 positional difference.
\label{PMvalues}
\end{table*}

\section{MCMC priors}

\begin{table*}[h!]
\centering
\caption{Priors considered for each free parameter.}
\begin{adjustbox}{width=\textwidth}
\begin{tabular}[h!]{ccccccccccc}
\hline
 Parameter & Eps ind A & HD 13931 & HD 115954 & HD 211847 & HD 219077 & HD 222155 & HIP 70849 \\ 
\hline
 \textit{a} (au) & [1,20] & [1,10] & [1,10] & [1,100] & [1,10] & [1,10] & [1,10] \\
 Eccentricity & [0,0.95] & [0,0.95] & [0,0.95] & [0,0.95] & [0,0.95] & [0,0.95] & [0,0.95] \\ 
 Inclination (°) & [0,180] & [0,180]  & [0,180] & [0,180] & [0,180] & [0,180] & [0,180] \\
 Mass (\Mjup) & [1,20] & [1,20] & [1,20] & [1,500] & [1,20] & [1,20] & [1,20] \\  
 $\Omega$ (°) & [0,360] & [0,360] & [0,360] & [0,360] & [0,360] & [0,360] & [0,360] \\ 
 $\omega$ (°) & [0,360] & [0,360] & [0,360] & [0,360] & [0,360] & [0,360] & [0,360] \\  
 Phase & [0,1] & [0,1] & [0,1] & [0,1] & [0,1] & [0,1] & [0,1] \\ 
 Jitter (m/s) & [0,10] & [0,10] & [0,10] & [0,20] & [0,10] & [0,20] & [0,10] \\
\hline
 Star mass (\Msun) & $0.76 \pm 0.04$ & $1.02 \pm 0.05$ & $1.18 \pm 0.06$ & $0.94 \pm 0.05$ & $1.15 \pm 0.06$ & $1.13 \pm 0.11$ & $0.63 \pm 0.03$ \\
 Distance (pc) & $3.622 \pm 0.004$ & $44.2 \pm 1.4$ & $218 \pm 2$ & $50.6 \pm 3.3$ & $29.3 \pm 0.2$ & $49.1 \pm 0.9$ & $24.0 \pm 0.7$ \\
 \hline
 & H03: [-41,-39] & Hir94: [-1,1] & ELODIE: [-15,-13] & C98: [5,7] & C98: [-31,-29] & ELODIE: [-5,-3] & H03: [-1,1] \\
 Instrumental & H15: [-41,-39] & Hir04: [-1,1] & SOPHIE: [-15,-13] & C07: [5,7] & C07: [-31,-29] & SOPHIE: [-5,-3] & H03: [-1,1] \\
 offset (km/s) & UVES: [-1,1] & & SOPHIE+: [-15,-13] & & H03: [-31,-29] & SOPHIE+: [-5,-3] & \\
 & LC: [-41,-39] & & & & AAT: [-1,1] & & \\
 & VLC: [-41,-39] & & & & & \\
\hline
\end{tabular}
\end{adjustbox}
\label{table_priors_planets}
\end{table*}

\newpage

\section{MCMC results}

\begin{figure}[h!]
  \centering
\includegraphics[width=1\textwidth]{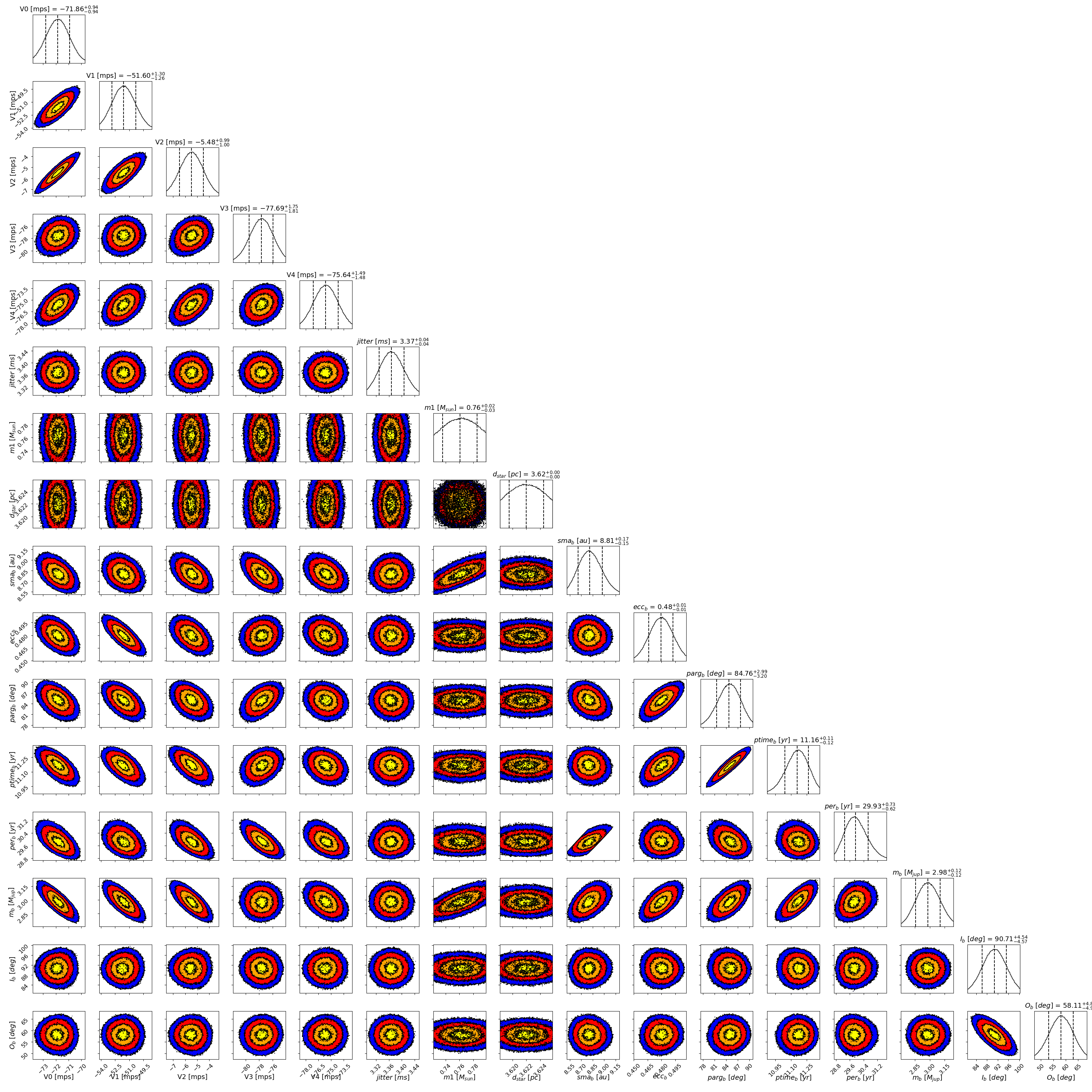}
\caption{Corner plot of the posteriors' fit of Epsilon Indi A combined RV and absolute astrometry. An offset of 39.9 km/s was added to V0, V1, V3, and V4 to improve readability.
\label{MCMC_EpsInd}} 
\end{figure}

\begin{figure}[t!]
  \centering
\includegraphics[width=1\textwidth]{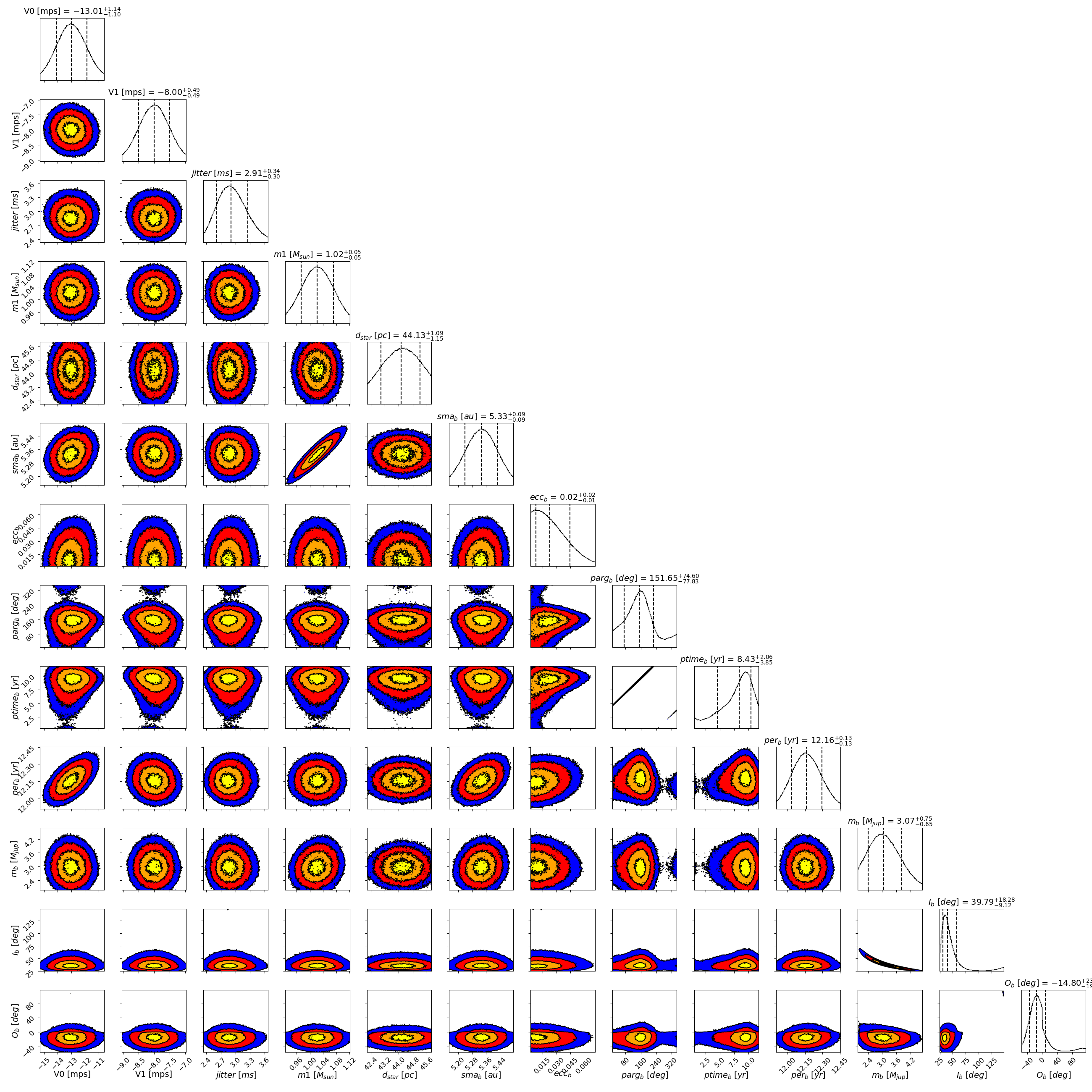}
\caption{Corner plot of the posteriors' fit of HD 13931 combined RV and absolute astrometry.
\label{MCMC_HD13931}} 
\end{figure}

\begin{figure}[t!]
  \centering
\includegraphics[width=1\textwidth]{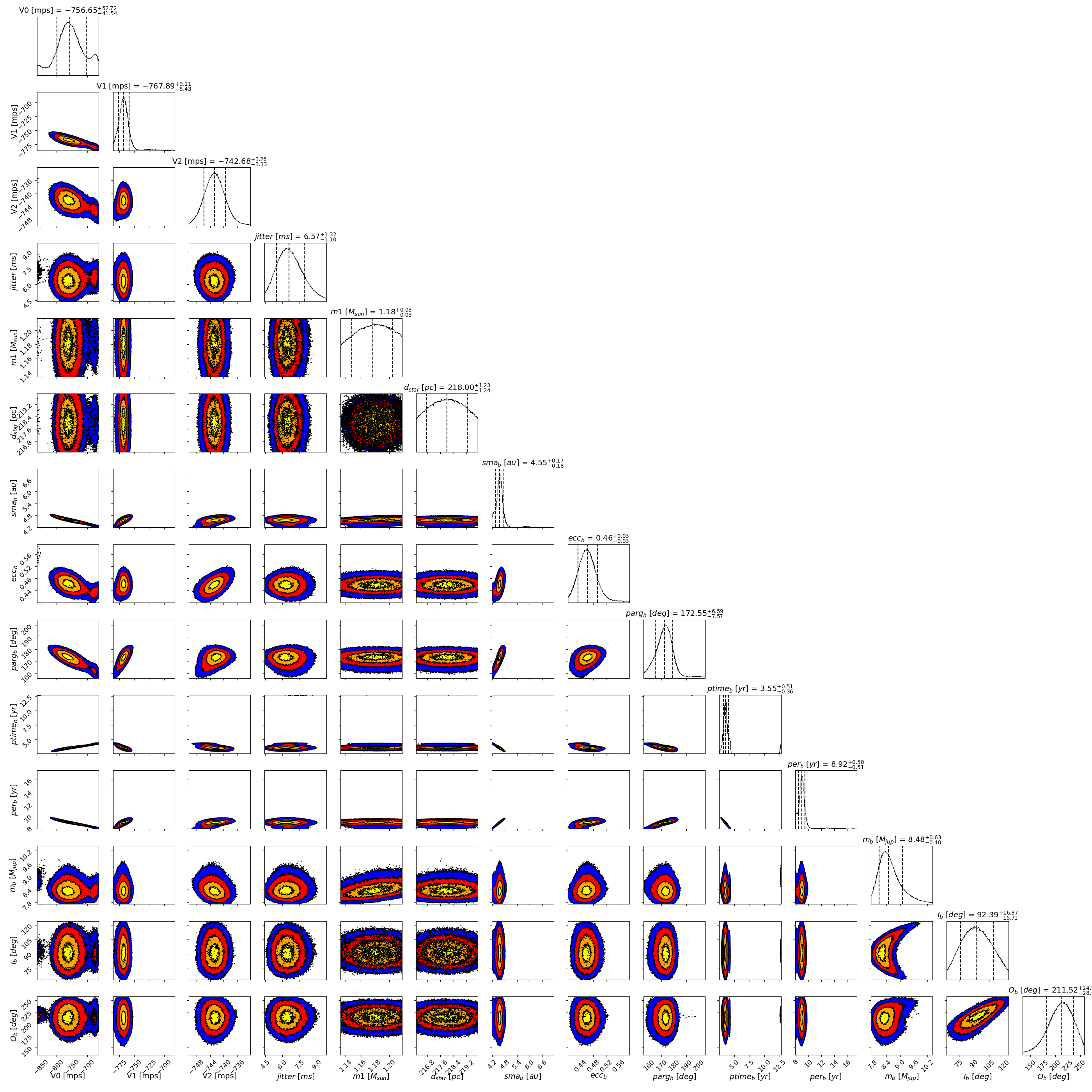}
\caption{Corner plot of the posteriors' fit of HD 115954 combined RV and absolute astrometry. An offset of 14 km/s was added to V0, V1, and V2 to improve readability.
\label{MCMC_HD115954}} 
\end{figure}

\begin{figure}[t!]
  \centering
\includegraphics[width=1\textwidth]{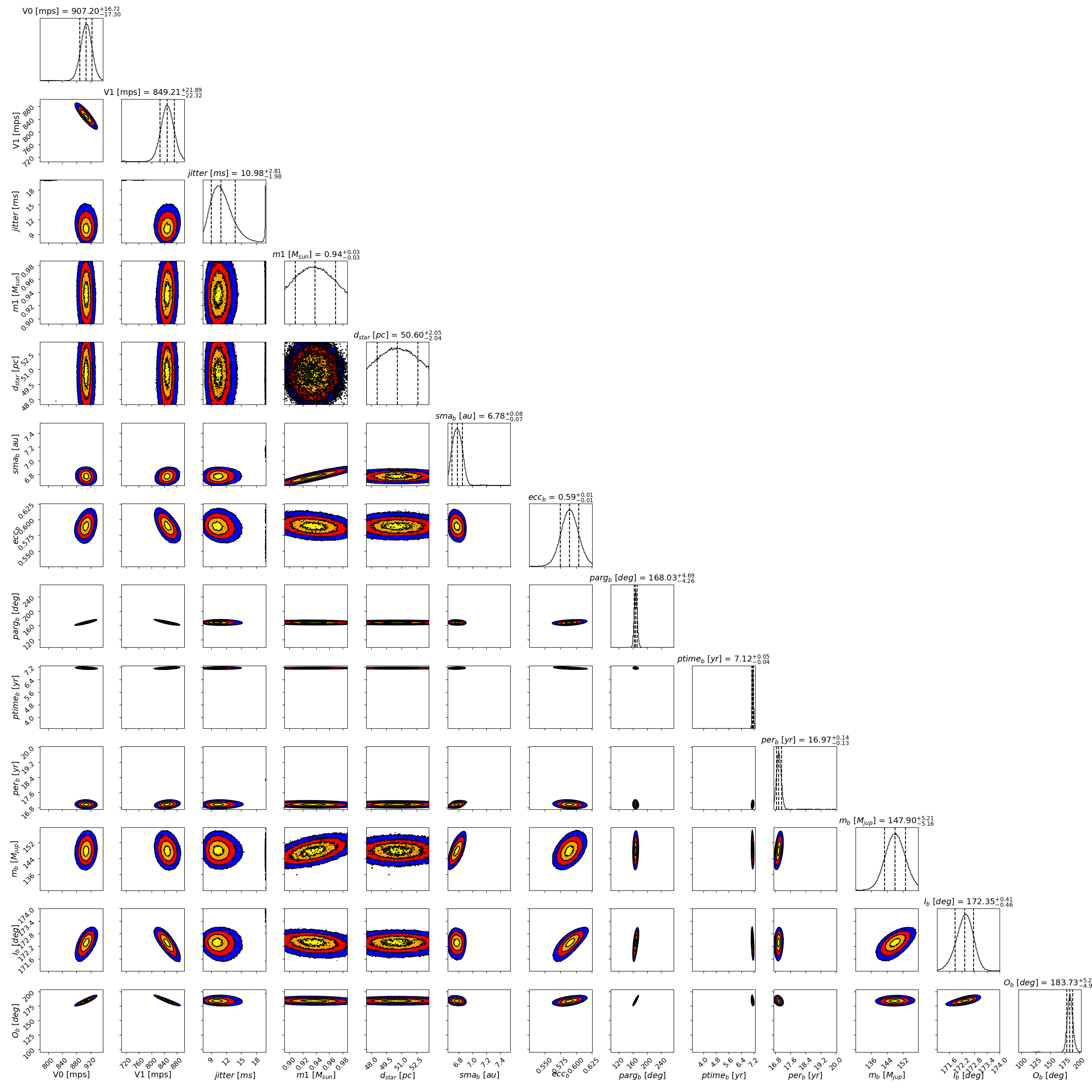}
\caption{Corner plot of the posteriors' fit of HD 211847 combined RV, relative astrometry, and absolute astrometry. An offset of 6 km/s was subtracted to V0 and V1 to improve readability.
\label{MCMC_HD211847}} 
\end{figure}

\begin{figure}[t!]
  \centering
\includegraphics[width=1\textwidth]{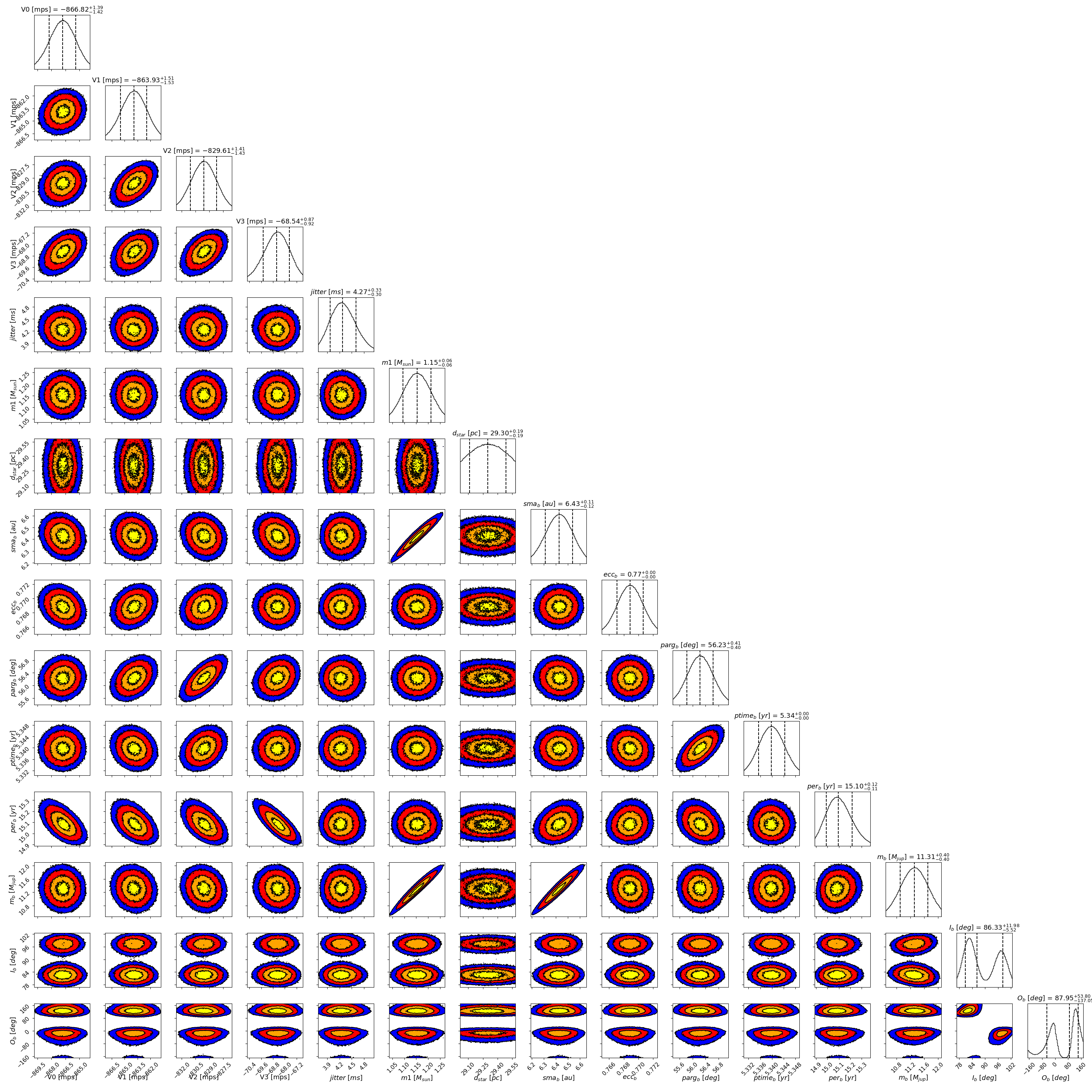}
\caption{Corner plot of the posteriors' fit of HD 219077 combined RV and absolute astrometry. An offset of 30 km/s was added to V0, V1, and V2 to improve readability.
\label{MCMC_HD219077}} 
\end{figure}

\begin{figure}[t!]
  \centering
\includegraphics[width=1\textwidth]{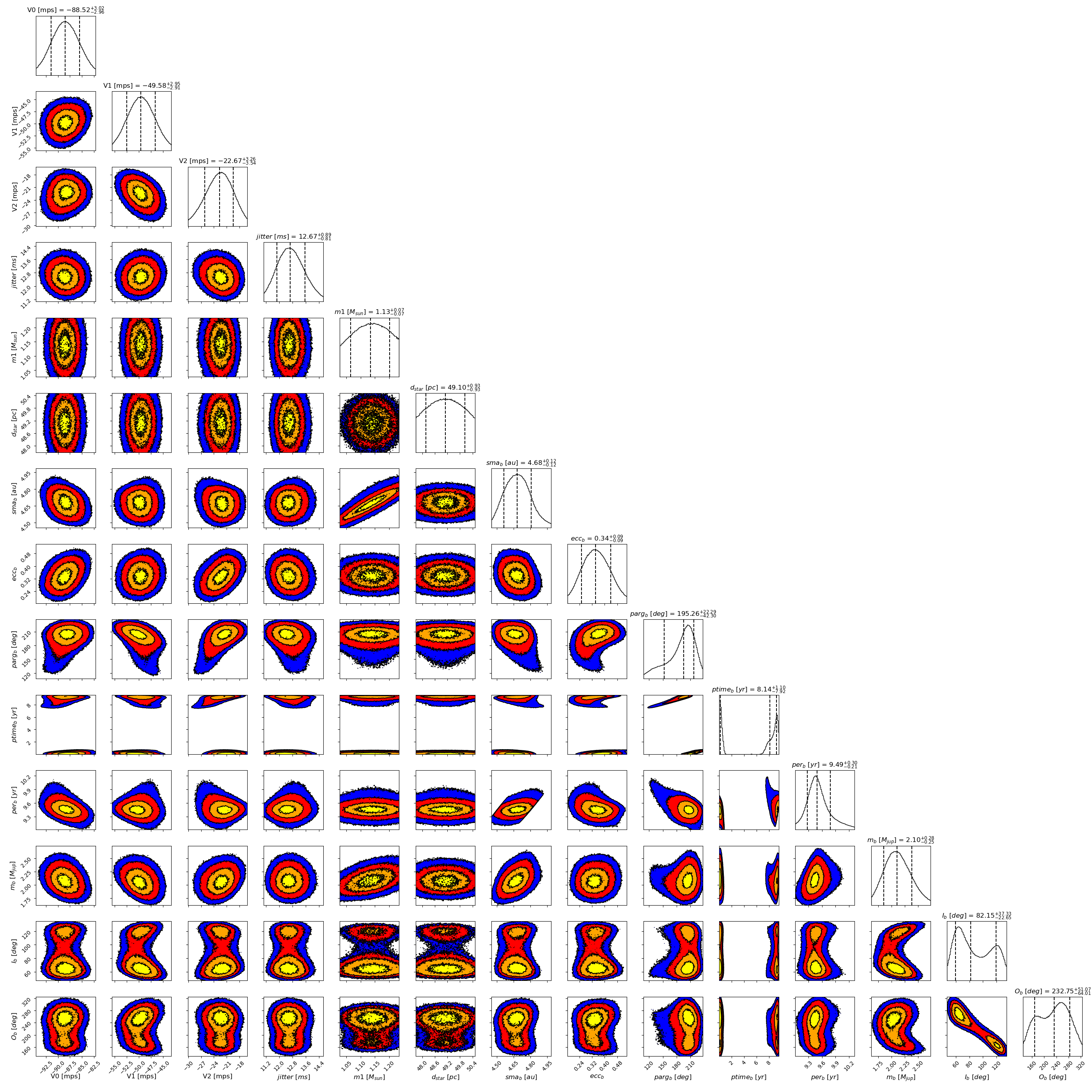}
\caption{Corner plot of the posteriors' fit of HD 222155 combined RV and absolute astrometry. An offset of 3.9 km/s was added to V0, V1, and V2 to improve readability.
\label{MCMC_HD222155}} 
\end{figure}

\begin{figure}[t!]
  \centering
\includegraphics[width=1\textwidth]{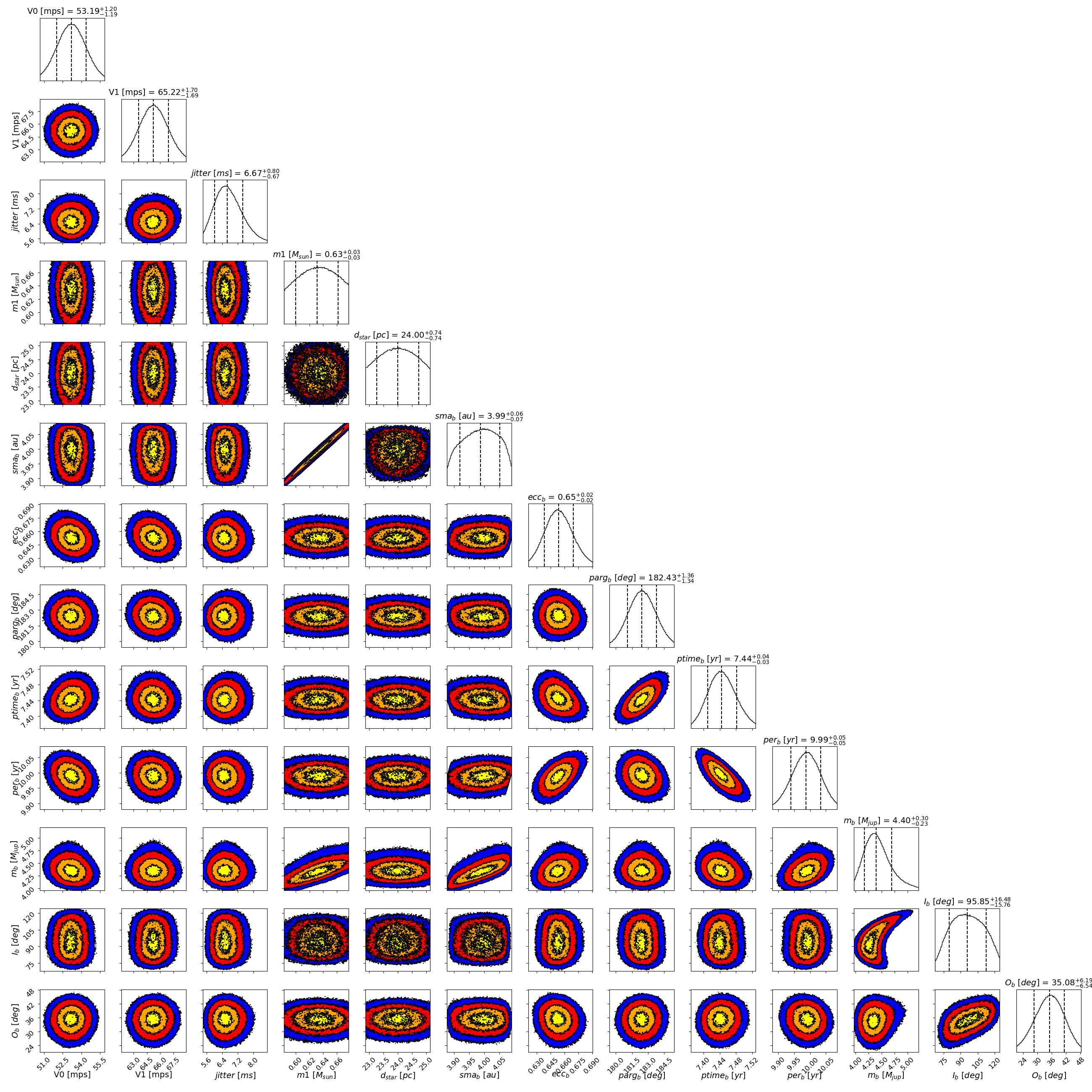}
\caption{Corner plot of posteriors fit of HIP 70849 combined RV and absolute astrometry.
\label{MCMC_HIP70849}} 
\end{figure}
\FloatBarrier
\end{appendix}
\end{document}